\def\figsize{9.5cm}
\documentstyle[prl,twocolumn,aps,rotate]{revtex}
\def \frontmatter {\twocolumn[\hsize\textwidth\columnwidth\hsize\csname@twocolumnfalse\endcsname}
  
\def\js12{{\cal C}}

\begin{document}
\draft

\input epsf.sty

\title{New Stopping Criteria  for Segmenting DNA Sequences}

\author{Wentian Li} 

\address{Laboratory of Statistical Genetics, Box 192\\
Rockefeller University, 1230 York Avenue, New York, NY 10021, USA.}

\date{\today}

\maketitle

\begin{abstract}

We propose a solution on the stopping criterion in segmenting
inhomogeneous DNA sequences with complex statistical patterns. 
This new stopping criterion is based on Bayesian Information 
Criterion (BIC) in the model selection framework.  When this
stopping criterion is applied to a left telomere sequence of yeast 
{\sl Saccharomyces cerevisiae} and the complete genome sequence 
of bacterium {\sl Escherichia coli}, borders of biologically meaningful 
units were identified (e.g. subtelomeric units, replication origin, 
and replication terminus), and a more reasonable number of domains 
was obtained. We also introduce a measure called segmentation 
strength which can be used to control the delineation of large domains. 
The relationship between the average domain size and the threshold 
of segmentation strength is determined for several genome sequences.
\end{abstract}

\bigskip

%\ifnum \tipo = 1
% 
%\narrowtext
%\fi

DNA sequences are usually not homogeneous. Regions with
high concentrations of G or C bases alternate with regions
which lack G or C \cite{bernardi95}; stretches of sequences 
with an abundance of CG dinucleotide (C$p$G island) interrupt 
regular sequences; coding regions distinguish themselves from 
non-coding regions by the strong periodicity-of-three pattern, etc.  
The alternation of long (e.g. $>$ 300 kilobases) G+C rich and 
G+C poor regions (also known as ``isochores" \cite{bernardi95}) 
is shown to be related to chromosome bands, gene density, and 
perhaps chromosomal structure \cite{bernardi95}. The concepts 
of inhomogeneity and domains can also be generalized recursively 
to different length scales, and such domains-within-domains 
phenomena have indeed been observed in DNA sequences \cite{seg,li97}. 
These hierarchical patterns are the cause of the fractional 
long-range correlations and 1/f spectra observed in DNA sequences 
\cite{lrc}. There have been discussions of the possible biological 
meaning of this hierarchical pattern \cite{audit} and its 
connection to other complex systems\cite{lrc2}. 

Computational methods used to identify homogeneous regions are called 
segmentation procedures \cite{seg,seg_other} which 
are important for many DNA sequence analysis tasks: detecting the 
existence of isochores, identifying complicated repeat patterns 
within telomeres and centromeres, determining coding-noncoding 
borders \cite{prl00}, etc. Segmentation procedures can also be
applied to any inhomogeneous/disorder media (e.g. 1-dim solid,
spinglass chain) or nonstationary time series (e.g. symbolic
dynamics) to determine the domain borders or turning points.
An application of the segmentation procedure to determine
the mobility edge of vibrational states in disordered materials
can be found in \cite{edge}.  The segmentation procedure and the 
physical fragmentation \cite{frag} are highly reminiscent of each 
other \cite{frag_ran}. The ease of a segmentation procedure
directly affects the scaling exponent of the size distribution
in a fragmentation \cite{frag_ran}.

In the segmentation procedure proposed in \cite{seg}, 
one crucial step -- the stopping criterion -- is arbitrarily determined. 
This is because this criterion is presented within 
the framework of hypothesis testing. It is common in this framework 
to reject or accept the null hypothesis based on a chosen significance 
level, typically, 0.01 or 0.001. Not choosing other levels, say, 
0.025 or $10^{-6}$, is to some extent arbitrary.  Another practical 
problem of the criterion in \cite{seg} is that it is extremely hard 
to halt the recursion at a large length scale even with a very 
small significance level, whereas 
many biologically interesting domains such as isochores are large.
We solve these problems here by discussing segmentation
in a new framework -- the model selection framework. As a result, 
an alternative meaning of segmentation is proposed, and a minimum 
requirement for choosing one model over another is introduced.

In the model selection framework,  basic 1-to-2 segmentation is carried
out as a comparison of two stochastic models of the DNA sequence: 
before the segmentation, the sequence is modeled by a homogeneous 
random sequence (with three base composition parameters); 
after the segmentation, by {\sl two} homogeneous 
random sequences separated by a partition point (with seven 
parameters). Whether a 1-to-2 segmentation should be continued or
not is determined by whether the two-random-subsequence model 
is better than the one-random-sequence model. In model selection, 
the answer to this question is determined by two factors: 
first, the model's ability to fit the data; and second, the model's 
complexity. Overfitting and underfitting models are not considered 
to be good, either because of high model complexity or because of 
poor fitting performance. The Bayesian information criterion (BIC) is
a proposal for balancing the two factors, defined as \cite{bic}:
\begin{eqnarray}
\label{eq_bic}
BIC  &=& -2\log(\hat{L}) + \log(N) K + O(1) +O(\frac{1}{\sqrt{N}}) 
+ O(\frac{1}{N}) \nonumber \\
&\approx & -2\log(\hat{L}) + \log(N) K
\end{eqnarray}
where $\hat{L}$ is the maximum likelihood \cite{edwards}, $K$ the 
number of parameters in the model, and $N$ the number of data points.
BIC is an approximation of the logarithm of integrated likelihood 
of a model multiplied by $-2$ \cite{bic}.
The integrated likelihood represents the overall performance of 
a model. The better the model, the larger the integrated likelihood, 
and thus the smaller the BIC. A similar concept
is the Akaike Information Criterion (AIC) \cite{aic_book}, 
with the $\log(N)$ term in Eq.(\ref{eq_bic}) replaced by 2. 
BIC penalizes complex models more severely than AIC.

We show here that the entropy-based segmentation in \cite{seg} 
can be recast in the likelihood framework \cite{edwards}, which in 
turn can be generalized to a model selection framework \cite{wli_recomb01}.
The likelihoods of the random-sequence model and the 
two-random-subsequence model (before and after a 1-to-2 
segmentation) are: $L_1(\{ p_\alpha \}) = \prod_{\alpha} p_\alpha^{N_\alpha}$,
$L_2(\{ p_\alpha^l\}, \{ p_\alpha^r\}, N_l) =
\prod_{\alpha}  (p_\alpha^l)^{N_\alpha^l} 
\prod_{\beta} (p_\beta^r)^{N_\beta^r} $, where 
$\{ p_\alpha \}, \{ p_\alpha^l\}, \{ p_\alpha^r\}$ ($\alpha$=A,C,G,T)
are the base composition parameters for the whole sequence, left 
and right subsequence, respectively; 
$\{ N_\alpha\}, \{ N_\alpha^l \}, \{ N_\alpha^r \}$ 
are the corresponding base counts; and $N_l$ is the size of the
left subsequence. The maximum likelihood estimation
of the parameters is simply $\hat{p_\alpha} = N_\alpha/N$, and the
maximum  log likelihoods before and after segmentation are
$\log\hat{L_1} = N E$ and 
$\log\hat{L_2} = N^l E^l+ N^r E^r$,
where $E, E^l, E^r$ are the entropies for the whole, left, and 
right sequences. The segmentation position $N_l$ is also
a parameter in the model, and is determined by the position 
that maximizes the likelihood (though this parameter is 
discrete and it's range changes with $N$). The increase 
of log-likelihood is 
$\log(\hat{L_2}/\hat{L_1}) = NE - (N^l E^l+ N^r E^r) = N \cdot \hat{D}_{JS}$,
where $\hat{D}_{JS}$ is the maximum of Jensen-Shannon divergence
$D_{JS} = E - (N^l E^l+ N^r E^r)/N$ \cite{lin91,seg}.

We require that the BIC be reduced by the segmentation for 
the procedure to continue, i.e.
$\Delta BIC <0 $,  
which leads to (note $K_2=7$ and $K_1=3$ \cite{note_df}):
\begin{equation}
\label{eq_bic_stop}
 2N \hat{D}_{JS} > 4 \log(N). 
\end{equation}
Eq.(\ref{eq_bic_stop}) is our new stopping criterion.

{\sl Lower (relaxed) bound of the significance level:}
The stopping criterion in Eq.(\ref{eq_bic_stop}) differs from 
the criterion in \cite{seg} in that the significance level 
cannot be arbitrarily relaxed. The criterion in \cite{seg}
compares the maximum $D_{JS}$ with that of a random sequence.
If the sequence is indeed random, $2N \hat{D}_{JS}$ is known to 
follow a $\chi^2$ distribution \cite{note_df}, and the tail-area
under this distribution is the corresponding significance level 
\cite{note_p}. The new criterion in Eq.(\ref{eq_bic_stop}) requires 
that the significance level cannot be too relaxed.  For example, if 
$N$ is 1 kilobase,  Eq.(\ref{eq_bic_stop}) is equivalent to setting
the significance level 1.48 $\times$ 10$^{-5}$, and if $N$ is 1 
megabases, it is  2.86 $\times$ 10$^{-11}$. The dependence 
of Eq.(\ref{eq_bic_stop}) on the sequence length $N$ has 
important practical implications: the stopping criterion 
in Eq.(\ref{eq_bic_stop}) is not fixed but adjustable. It is 
particularly important for a long sequence, when the 
criterion in \cite{seg} may not be able to stop segmentations
with large $2N \hat{D}_{JS}$.

\def\figsize{10.0cm}
\begin{figure}
\centerline{
  \epsfysize=\figsize {\rotate[r]{\epsfbox{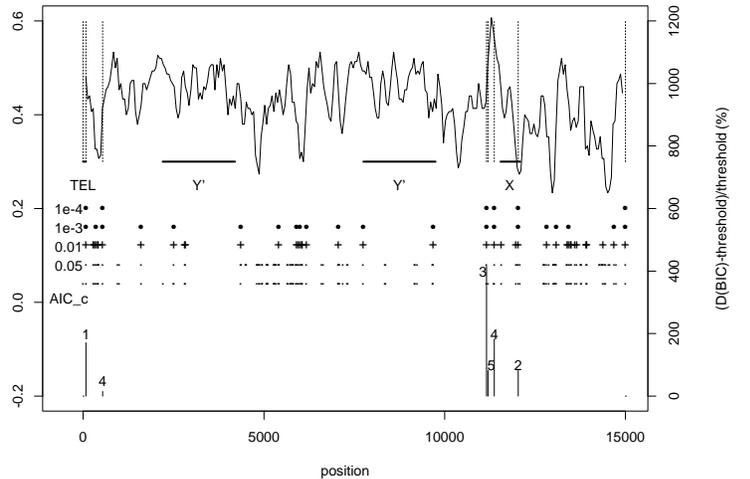}}}
}  
\caption{
Partition points  determined by the segmentation with the stopping
criterion Eq.(\ref{eq_bic_stop}) for the left telomere of yeast 
{\sl S. cerevisiae} chromosome 12 (dashed vertical lines). 
The partition points determined by AIC (dot) (with the high-order 
term included), hypothesis testing framework with significance 
level of 0.05 (dot), 0.01 (cross), 0.001 and  0.0001 (solid dot) are 
shown for comparison. Also shown is the G+C content in moving windows 
(window size=150 bases, moving distance=51 bases). The location 
of the telomeric sequence (TEL) and subtelomeric sequences (Y'
and X) are marked.  The lower plot shows the segmentation strength 
$s$ of a 1-to-2 segmentation.  The numbers are the order in which 
the segmentation is carried out.
}
\label{fig1}
\end{figure}

In Fig.\ref{fig1}, we illustrate the new criterion for the left telomere 
of chromosome 12 of  yeast {\sl Saccharomyces cerevisiae} 
\cite{yeast12}. It is known that telomere sequences are compositionally 
complex. There is a highly repetitive sequence called TEL at the tip 
of the telomere (for yeast, it is 5'-C$_{1-3}$A-3'). There are also 
subsequences that are conserved among different yeast chromosomes: 
the Y' and X subtelomeric sequence \cite{yeast}. A segmentation 
procedure can be applied to telomere sequences to identify some 
compositionally distinct elements \cite{wli_tel}.  It can be seen from 
Fig.\ref{fig1} that the criterion in Eq.(\ref{eq_bic_stop}) manages
to delineate the borders for TEL and X elements \cite{note_elem}. 
Although Eq.(\ref{eq_bic_stop}) missed the two Y' elements,
an indication that Y' elements are not compositionally distinct,
it is the cost of avoiding many false positives.

{\sl Segmentation strength:} Although a lower (relaxed) bound of 
the significance level is set in Eq.(\ref{eq_bic_stop}), no limit 
on the upper (stringent) bound is possible. We introduce a measure 
for segmentation strength $s$ \cite{wli_recomb01}:
\begin{equation}
\label{eq_s}
 s = \frac{2N \hat{D}_{JS} - 4 \log(N)} {4 \log(N)} ,
\end{equation}
and the stringency level can be raised by choosing a non-zero value of 
the threshold $s_0$: $ s > s_0 > 0$. Eq.(\ref{eq_bic_stop}) is equivalent 
to $s_0=0$.  The prominence of TEL and X elements is indicated by their 
large segmentation strength  ($s=$ 170.66\%, 84.6\%, and 416.33\%; see
the lower plot of Fig.\ref{fig1}).  These segmentations are also chosen 
earlier in the recursive segmentation (being first, second, and third).

\def\figsize{9.0cm}
\begin{figure}
\centerline{
  \epsfysize=\figsize {\rotate[r]{\epsfbox{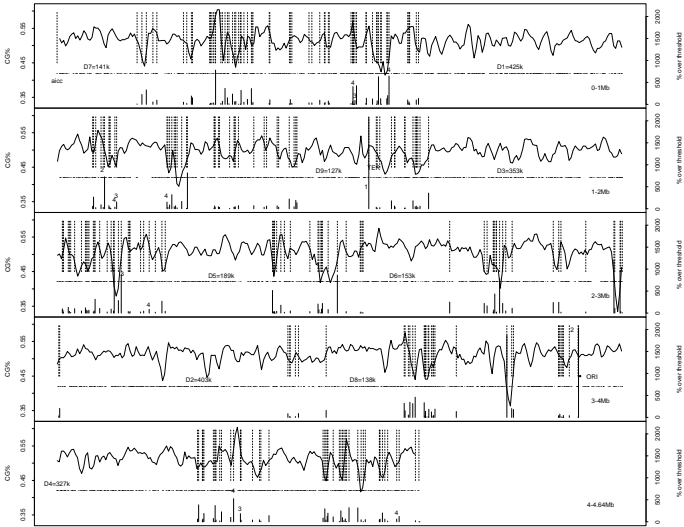}}}
}  
\caption{
Segmentation points determined by Eq.(\ref{eq_bic_stop}) for 
{\sl E. coli} genome (dashed vertical lines). Also shown are 
the G+C content in moving windows (window size=9000 bases,
moving distance=3571 bases), and the segmentation strength $s$.
The segmentation points determined by the AIC-based stopping
criterion are shown  by the dots.  The replication 
origin, replication terminus, and the 9 largest domains are marked in
the plot. Each one of the subplots represents 1 megabase of 
the sequence (total length is 4.639 megabases).
}
\label{fig2}
\end{figure}

{\sl Minimum Domain Size:} To test a model on a dataset, the number
of samples must be larger than the number of parameters in the 
model. Since we compare two models with 3 and 7 parameters,
respectively, the sequence has to contain at least 7 bases before
the segmentation, and 3 bases after the segmentation.  Unlike
the criterion in \cite{seg}, these minimum size requirements
are not set arbitrarily.

{\sl Binary and 12-Symbol Sequences:} For many practical
applications of the segmentation procedures, DNA texts are converted 
to symbolic sequences with less (or more) than four symbols. For 
example, the two-symbol sequence with symbols S (for strong, G and C) 
and W (for weak, A and T), is frequently used for studying large-scale
homogeneous domains.  The stopping criteria for binary sequences
can be modified easily: with $K_1=1$ and $K_2=3$, the right-hand-side
of  Eq.(\ref{eq_bic_stop}) becomes 2$\log$(N).  For coding 
region recognition, it is proposed in \cite{prl00} that a DNA 
sequence can be converted to a 12-symbol sequence: each symbol
contains information on both the base and the codon position
(i.e. $A_1,C_1, G_1, T_1, A_2, \cdots$). With $K_1=9$ and $K_2=19$,
the  stopping criteria in Eq.(\ref{eq_bic_stop}) become 
$ 2N \hat{D}_{JS} >  10\log$(N).

{\sl Threshold for segmentation strength and domain sizes:} Since 
Eq.(\ref{eq_bic_stop}) does not provide an upper (stringent) limit 
on the significance level, there is still some degree of subjectivity 
in our segmentation procedure. If one is interested in largest domains, 
or the strongest segmentation signals, the threshold for segmentation 
strength $s_0$ should be set larger than zero. Taking the complete 
sequence of {\sl Escherichia coli} genome \cite{ecoli} for example, 
the replication origin and the replication terminus presents the two 
most significant segmentation signals. If the $s_0$ is set to 20, 
only these two 1-to-2 segmentations will make the cut.

The larger the $s_0$, the larger the domain sizes in the final 
configuration. The relationship between the two is empirically 
determined by segmentations on several genome sequences, shown in 
Fig.\ref{fig3}. It can be seen that the relationship is not universal 
for all sequences: with the same $s_0$, sequences with high 
compositional complexity (e.g. MHC sequence) contain smaller 
domain sizes in the final configuration than sequences with lower
complexity (e.g. yeast).  It can also be seen that in order to 
reach the average size of isochore (300 kilobases), $s_0$ should 
be set as large as 500\%. 

\def\figsize{9.0cm}
\begin{figure}
\centerline{
  \epsfysize=\figsize {\rotate[r]{\epsfbox{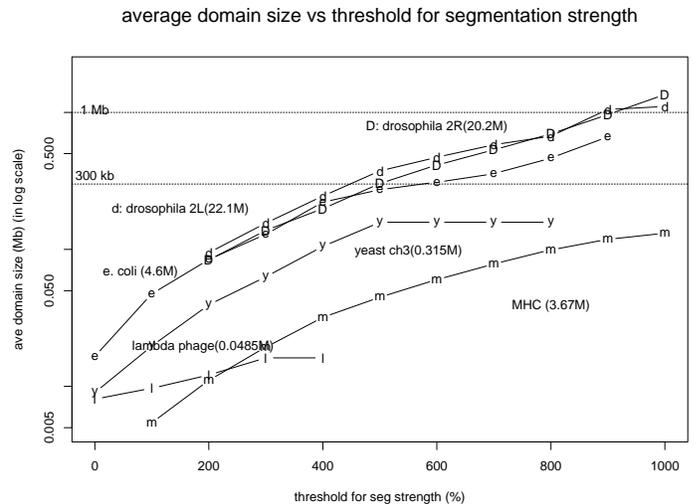}}}
}  
\caption{
Average domain size vs. segmentation strength $s_0$ for these sequences:
human major histocompatibility complex (MHC), $\lambda$ bacteriophage, 
chromosome 3 of {\sl S. cerevisiae}, {\sl E. coli}, left and right 
arms of chromosome 2 of {\sl Drosophila melanogaster}.
}
\label{fig3}
\end{figure}

{\sl Domain size distribution}: Another indirect evidence that 
our new stopping criterion is more reasonable than the one in 
\cite{seg} (with a typical significance level) can be seen by 
examining the domain size distribution in the final configuration. 
The 281  domains in the {\sl Escherichia coli} genome in 
Fig.\ref{fig3} are ranked by size. These sizes are 
plotted against the rank (Zipf's plot) in Fig.\ref{fig4}. 
The Zipf's plot for sizes from rank 4 to rank 180 approximately
exhibit a power-law $1/r^{1.21}$ (Fig.\ref{fig4}). This is similar 
to the power-law behavior in Zipf's plot of many other natural and 
social phenomena (known as Zipf's law \cite{zipf,zipf_web} when the
scaling exponent is close to  $-1$).

When a more relaxed stopping criterion is used, there is a lack
of large domains. We illustrate this by a AIC-based segmentation
which is equivalent to the criterion in \cite{seg} with the
significance level of 0.091578.  The Zipf's plot for domains derived
from the AIC-based segmentation is not a power-law function.
Even a forced curve-fitting by a power-law function leads 
to a slope merely $\sim -0.5$. This indicates that the size 
distribution by criterion Eq.(\ref{eq_bic_stop}) is more self-similar, 
more balanced between the small and large domains than those
by the AIC-based segmentation.

\def\figsize{7.0cm}
\begin{figure}
\centerline{
  \epsfysize=\figsize {\rotate[r]{\epsfbox{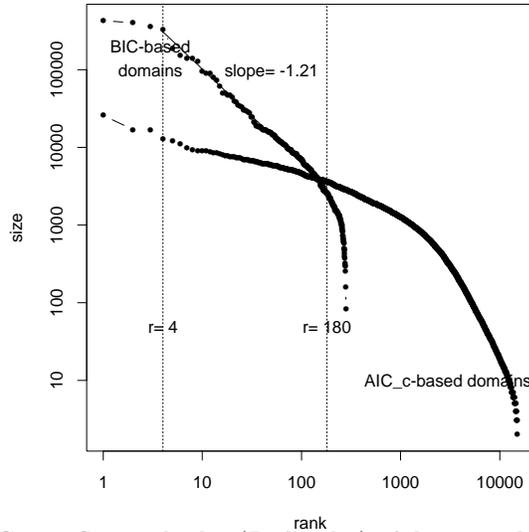}}}
}  
\caption{
Size-rank plot (Zipf's plot) of domains obtained by segmentation with
the stopping criterion in Eq.(\ref{eq_bic_stop}). Those obtained 
by the AIC-based segmentation are also shown.
}
\label{fig4}
\end{figure}

In summary, this paper solves a problem encountered in \cite{seg}
that recursive segmentation is not easy to stop even when 
a stringent significance level is used (the most stringent 
significance level in the SEGMENT program \cite{seg_prog} 
is $10^{-6}$). This solution allows us to investigate much 
larger domains and longer-range hierarchical correlation in 
DNA sequences. The framework from which our solution is 
derived is also ideal for generalizations to other more 
complicated situations.  Determining the number of domains
in a DNA sequence, like any other descriptions of the sequence,
is relative -- it is relative to the length scale of interests, 
relative to the model used. By changing the segmentation strength,
we essentially change the level of description of the sequence.

The work is supported by the grant K01HG00024 from NIH. I thank 
J\'{o}se Oliver for sending me the partition points used in 
Fig.\ref{fig1}, produced by the SEGMENT program \cite{seg_prog}.
This paper is dedicated to XML.


\begin{references}
\bibitem{note_df}
In \cite{seg,prl00}, the partition point is not considered as a
free parameter so $K_2=6$ and $\chi^2_{df=6-3}$ is used. The maximization
over the partition point is discussed separately \cite{prl00}. It
is more convenient to consider the partition point as a free parameter
and $K_2=7$. Consequently, $\chi^2_{df=7-3}$ is used.
\bibitem{note_p}
The term significance level used here is consistent with
the typical usage in statistics literature (e.g. s=0.01).
Note that in \cite{seg}, $1-s$ is called the significance level.
\end{references}
\end{document}